# The manifestation of the band structure in the photon emission spectrum of the fast above-barrier oriented particle

E.A. Mazur [a]

[a] National Research Nuclear University "MEPHI", Kashirskoe sh.31, Moscow, 115409, Russia

The probability of emission of photons in non-dipole case by channeled particle is calculated. The emission of hard photons with an energy comparable to the energy of the incident channeled particle is examined. The calculation of the quasi-Bloch energy spectrum of the oriented fast charged particle entering the crystal at an angle substantially greater than the Lindhard angle is performed. The initial and final spectra of the channeled particle belonging to a different set of band wave functions corresponding to different energies are used. The processes of the photon generation by the quantum crystal-oriented particle entering into the crystal at an angle both greater and smaller than the Lindhard angle are considered on the equal footing. It is shown that the spectrum of hard photon emission consists of a set of the well-observed emission lines. The probability of the nondipole photon radiation by the channeled particle with the energy $E = 10 GeV$ within a small solid angle oriented along the propagation direction of the channeling particle is shown to be equal to $w_{if}(\omega_{if}) = 0.2$ s$^{-1}$. It has been shown that the nondipole processes of the radiation of hard photons by the channeled particles are well observed experimentally.

## 1. Equation for the wave function of the fast charged particle in the dimensionless form

The theory of the photon radiation by the oriented fast charged particles has been built in a number of works [1-9]. At the same time, however, the questions of nondipole radiation of hard photons in the conditions when both the initial and final state of the channeled particle (CP) belong to the different bases, corresponding to the different total CP energies, have not been studied in detail. The fact that the matrix elements of the CP quantum transition are usually complex functions of the momentum transfer has not been taken into account in the calculations as well. Usually only the matrix elements of the first kind between the wave functions of the transverse motion of the CP $I^{(1)} = \langle \psi_n \exp(ikx) \psi_{n'} \rangle$ have been taken into account in the calculation. The role of the CP quantum transition matrix elements of the second kind $I^{(2)} = \langle \psi_n \exp(ikx) \partial/\partial x \, \psi_{n'} \rangle$, where both the wave function of the initial state and the derivative of the wave function of the final state act as the plates [1-9] has been little studied. The calculations are usually carried out for the CP sub-barrier transitions. In the present study, we investigated the role of non-dipole processes in the emission of hard photons with the consideration of above-barrier and sub-barrier CP transitions on an equal footing. The equation for the wave function of the relativistic oriented particle interacting with the crystal has the following form

$$\left[ E_{n,p}^2/c^2 - 2eE_{n,\vec{p}}U(\vec{r})/c^2 + e^2U^2(\vec{r})/c^2 + \hbar^2\Delta_{\vec{r}} - m^2c^2 + ie\hbar\vec{\alpha}\nabla_{\vec{r}}U(\vec{r})/c \right]\psi_{n,\vec{p}}(\vec{r}) = 0. \quad (1)$$

In this formula $\vec{\alpha} = \gamma^0\vec{\gamma}$, $\gamma^\mu (\mu=1,2,3)$ so that (1) for the unperturbed wave function $\psi_{n,\vec{p}}(\vec{r})$ of the fast oriented or channeled particle in the state with the quantum numbers $n, \vec{p}$ can be can be rewritten as:

$$\left( \frac{\hbar^2}{2m}\Delta_{\vec{r}} + \frac{E_{n,\vec{p}}^2 - m^2c^4}{2mc^2} \right)\psi_{n,\vec{p}}(\vec{r}) = \left( eE_{n,\vec{p}}U(\vec{r})/mc^2 \right)\psi_{n,\vec{p}}(\vec{r}). \quad (2)$$

Denote $E_{n,\vec{p}}/mc^2 = \gamma$, $eU(x) = V(x)$. As is known, if the point group of the crystal contains an inversion, then the relation $V_{-\vec{G}} = V_{\vec{G}}$ for the crystal potential appears to be true. For these crystals, such as silicon, germanium, InSb, GaAs the crystal potential should be written down in the following form:

$$V(x) = \sum_{G_x} V_{G_x} \exp(iG_x x) = V_0 + 2\sum_{G_x>0} V_{G_x} \cos(G_x x). \quad (3)$$

Here $G_x \equiv G$ is the reciprocal lattice vector in the direction perpendicular to the plane of the channeling. Take the concrete lattice potential with the approximate parameters corresponding to 110 silicon crystal plane at the channeling direction, resulting in the channeling equidistant planes. A significant part of the spatially inhomogeneous terms in the average potential of the planar crystal (3) are equal to zero in such crystals, and the remaining terms are expressed in terms of the Fourier components $V_{4n,0,0} (n=1,2,3...)$ of the spatial lattice potential $V(\vec{r})$, so that

$$V(x) = \bar{V} + 2\sum_{n=1,2} V_{4n,0,0} \cos(4Gx). \quad (4)$$

In other words, for the reasons of the crystal potential symmetry the specific direction of channeling in a silicon crystal is chosen, at which the average potential of the crystal planes is a set of of equidistant crystal planes. Limiting ourselves to the first two largest terms in (4), consider the problem of radiation of hard photons by electron and positron in the channeling. Choose $V_0 = -2V_G$ to select the zero value of the potential energy. Select here the dimensionless units. Let $2S = G_x x$ here, where $G_x$ is responsible for the first non-vanishing contribution to the potential. Redenote $\psi_{n,\kappa}(S) \equiv U(S)$. We obtain, renaming, $\psi_{n,\kappa}(S) \equiv U(S)$, $V(S) = V_0 + 2V_{G_x} \cos(2S)$ the following equation:

$$\left( \Delta_S - \frac{8E_{n,\bar{p}}}{c^2\hbar^2 G_x^2} V_0 + \frac{8m}{\hbar^2 G_x^2} \varepsilon_{n,\kappa} - \frac{8E_{n,\bar{p}}}{c^2\hbar^2 G_x^2} 2V_{G_x} \cos(2S) \right) U(S) = 0. \tag{5}$$

Here $G_x = nG$ is the first reciprocal lattice vector, whose potential is not zero. Comparing (5) with a known Mathieu equation with the dimensionless coefficients [10]

$$\frac{\partial^2 U}{\partial S^2} + (\tilde{a} - 2q\cos 2S)U(S) = 0,$$

we get:

$$\tilde{a} = \frac{8m\varepsilon_{n,\kappa}}{G_x^2 \hbar^2} - \frac{8EV_0}{\hbar^2 c^2 G_x^2}; \quad q = \frac{8EV_{G_x}}{\hbar^2 c^2 G_x^2}; \quad S = G_x x/2; \quad G_x \equiv nG_{\text{мин}} = nG = n2\pi/d; \quad E_\perp \equiv \varepsilon_{n,\kappa}$$

$V_0 = -2V_{G_x}$, $\tilde{a} = \frac{8m\varepsilon_{n,\kappa}}{n^2 G^2 \hbar^2} + \frac{16EV_{G_x}}{\hbar^2 n^2 c^2 G^2}$; $q = \frac{8EV_{G_x}}{\hbar^2 c^2 n^2 G^2}$. $\tilde{a} = \frac{\varepsilon_{n,\kappa}}{12.5eV} \frac{16}{n^2} - 2.4\gamma \frac{16}{n^2}$; $q = 2.4\gamma \frac{8}{n^2}$. For

Si $G_x \equiv 4G$, $n = 4$, $\tilde{a} = \frac{m\varepsilon_{n,\kappa}}{2G^2 \hbar^2} + \frac{EV_{G_x}}{\hbar^2 c^2 G^2}$; $q = \frac{EV_{G_x}}{2\hbar^2 c^2 G^2}$. Estimate the following terms for the electron: $\frac{2\hbar^2 G^2}{m} = \frac{2 \cdot 10^{-68} 10^{-20}}{10^{-30}} = 2 \cdot 10^{-18} J = 12.5 \text{ eV}$, $\frac{EV_G}{\hbar^2 c^2 G^2} = 2\gamma V_G / \frac{2\hbar^2 G^2}{m} \approx$

$\gamma 30 eV / 12.5 eV \approx 2.4\gamma$ Finally for Si $\varepsilon_{n,\kappa} = \tilde{a} - 2.4\gamma$, $q = 1.2\gamma$; where $\varepsilon_{n,\kappa}$ is measured in units of 12.5 eV. Subbarrier state are such that $\varepsilon_{n,\kappa} < 0$ as the full potential of the crystal with the selection of the zero energy has been dropped into negative values. For channeled positively charged particles $q > 0$, since $\bar{V}$ and $V_{400}$ are larger than zero. The value of $a$ can be either smaller or greater than zero, depending on the transverse energy of the channeled particle. Estimate the value of $q$ in equation (5). At 28 MeV, $\bar{V} \approx 15$ eV (ie the depth of the well equals 30 eV, $G_x = 10^{10}$ m$^{-1}$), we obtain $q \approx 11.2$, the value of $a$ varies depending on $E_\perp(P_x)$, and can be both large and small. The solutions of (5) have Bloch band character. At the same time the boundaries of the energy bands are determined by the well-known in the theory of Mathieu functions [10] numbers $a_r - q$, and $b_{r+1} - q$, where $r = 0, 1, 2...$. The allowed values of the transverse CP energy $E_\perp(P_x)$ are determined by the condition $a_r - q < a < b_{r+1} - q$, where $r$ is the number of the allowed band. The reckoning of certain bands to the discrete or continuous spectrum is conditional and is determined solely by the width of the CP band. The wave functions which represent the solution of equation (5) have the Bloch form:

$$u_{E_\perp}^{(n)}(x) = (N_x a_x)^{-0.5} \exp(iv_n(E_\perp)Gx) D_{E_\perp}^{(n)}(2Gx), \quad (6)$$

where the modulator of the band wave function $D(2Gx)$ is periodic in $x$ with the period $a_x/4$ equal to one-fourth (see (4) for the explanation of this fact) of the crystal lattice period along $x$-direction, $N_x$ is the the number of crystallographic planes of the crystal. $D(2Gx)$ obeys the following relation $D(2G(x+a_x/4)) = D(2Gx)$. The dimensionless Bloch modulators are subject the following normalization condition: $\int_{-a_x/8}^{a_x/8} \frac{dx}{a_x} D_{E_\perp}^{(n)}(2Gx) D_{E_\perp}^{(n')}(2Gx) = \delta_{nn'}$. Functions (6) are investigated in the theory of the Mathieu functions, where they are referred to as the the solutions of the Mathieu equations in the Floquet form [10]. When quasi momenta CP correspond to the boundaries of the energy bands of the CP, ie $G_n = 2\pi n/a$ $(n=0,1,2...)$, Floquet functions become the well-studied Mathieu functions $Cl_r(S,q)$ and $Sl_r(S,q)$ where with the symbol $Cl_r(S,q)$ the even CP wave functions corresponding to the lower edge of the energy band have been denoted, while $Sl_r(S,q)$ correspond to the odd wave functions of the CP related with the upper edge of the respective energy band of the transverse CP motion.

**2. Band structure and wave functions of the fast oriented electron and positron.**

Calculation of the band structure for a realistic potential U (x), resulting from the averaging potentials of the atoms that make up the crystal plane [1-9], is the task of the numerical calculation. Fig. 1 shows the results of this calculation for positron with the energy 28MeV in the channel corresponding to the crystallographic (110) plane in silicon Si. The band structure for the 15 zones for the electrons with the energies corresponding to the value $-300 < q < -10$ of the parameter $q$ is shown in Fig.1.

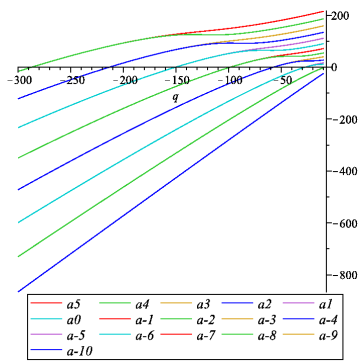

Fig. 1. Band structure for the oriented electron in a silicon crystal. Band number is numbered from negative number n= - 10 to positive number n=+5 with the notation a-n. Some band lines are merging with increasing modulus of q.

The wave functions as well as the density of the probability of finding a positron in the channel are shown in figs. 2-4.

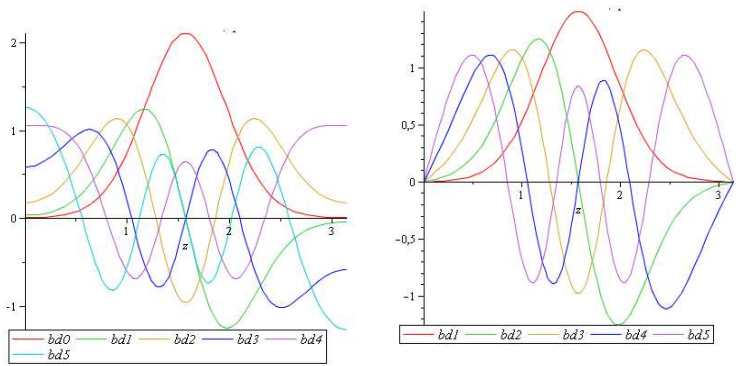

Fig.2. First six a.even and b. odd wave functions of the positron with the energy corresponding to the parameter q = 11.3 (about 50 MeV). The part of the wave functions corresponding to only the right half of the corresponding channel is shown. Designation bdn denotes band numbers in ascending order of band number from 0 to 5 for the even zones and from 1 to 5 for the odd bands. The value of the dimensionless variable $z = 2Gx = 4\pi x / a_x$ is plotted by axis z.

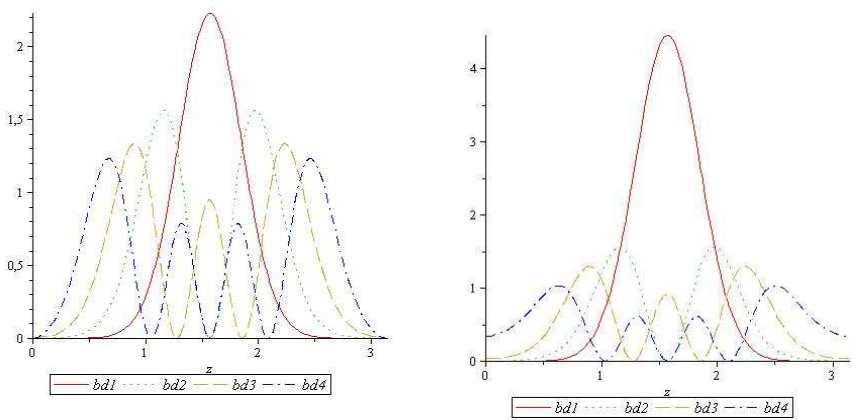

Fig. 3. Square module of the even (a) and odd (b) wave functions of positrons with an energy of 28 MeV in the planar channel (110) in a single crystal Si. (bd1) - first deep sub-barrier level; (bd2) - the second level in the middle of the channel; (bd3) - the first above- barrier band (level) ); (bd4) - second above-barrier band. The value of the dimensionless variable $z = 2Gx = 4\pi x / a_x$ is plotted by axis z.

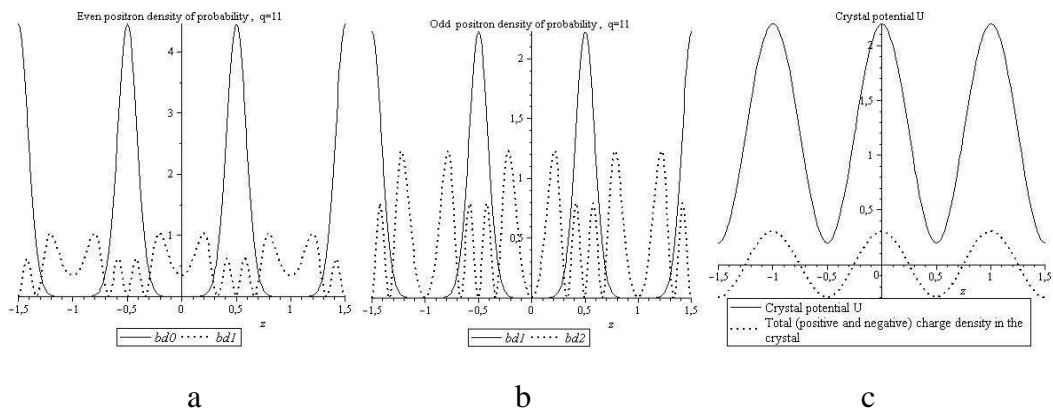

a　　　　　　　　　　b　　　　　　　　　　c

Fig. 4. Squared module of the even (a) and odd (b) wave functions of positrons with an energy of 28 MeV in the planar channel (110) in a periodic potential with parameters corresponding to the plane 110 in the silicon single crystal. x-coordinate is listed in dimensionless units as a fraction of the interplanar distance. a: (bd0) - the first deep sub-barrier zone; (bd1) - third near-barrier zone; b: (bd1) - the second zone; (bd2) - fourth (above-barrier) band. c: the graphs show in the same scale along the x-axis the planar potential U (solid line) and the total density distribution of the positive and negative charges in the crystal (line consisting of points).

The peaks of the density of the probability of finding a positron for high near-barrier zones are shifted to the edges of the channel, ie to the atomic planes.

**3. Emission of hard photons by the fast charged particle. Energy and pulse conservation laws.**

Describe the emission of photons by the channeled particles using the well known formalism [1-9]. For the differential probability of the photon emission write the following expression (all the notations are standard):

$$\frac{d^2w}{d\omega d\Omega} = \frac{e^2\omega}{2\pi}\sum_f |M_{if}|^2 \delta\left(\omega - \omega_{if} - \left(E_p^{\parallel} - E_{p-k}^{\parallel}\right)\right) \qquad (9)$$

In (9) the CP quantum transition matrix elements of both the first kind and the second kind [1-9] have been taken into account.

$$E_p^{\parallel} - E_{p-k}^{\parallel} = \omega - \frac{\omega}{2(E_i - \omega)}\left[\left(\theta^2 + E_i^{-2}\right)E_i - \omega\theta^2 \cos^2\varphi\right] \qquad (10)$$

Let us turn to the energy and momentum conservation laws. We get then

$$\delta\left(\omega - \omega_{if} - \left(E_p^{\parallel} - E_{p-k}^{\parallel}\right)\right) = \delta\left(\varepsilon_i - \varepsilon_f - \frac{\omega}{2(E-\omega)}\left[\left(\theta^2 + \frac{1}{\gamma^2}\right)E - \omega\theta^2 \cos^2(\varphi)\right]\right). \qquad (11)$$

At $\theta = 0$ obtain $\delta\left(\varepsilon_i - \varepsilon_f - \frac{\omega}{2(E-\omega)}\left[\left(\frac{1}{\gamma^2}\right)E\right]\right) =$

$$= \delta\left(A_i(E) - A_f(E-\omega) - 2q + 2q' - \frac{\omega}{2(E-\omega)}\left[\left(\frac{1}{\gamma^2}\right)E\right]\right), \quad q = \frac{8EV_{G_x}}{\hbar^2 c^2 n^2 G^2}; \quad q' = \frac{2(E-\omega)V_0}{\hbar^2 c^2 G_x^2} =$$

$= q(1 - \omega/E)$, $\omega/E = 1 - q'/q$. Denote $\omega/E = x$. Then $q' = q(1-x)$, $\frac{\omega}{2(E-\omega)} = \frac{1}{2}\frac{x}{(1-x)}$.

The delta function can be rewritten as:

$$\delta\left(12.5 A_i(q) - 12.5 A_f(q(1-x)) - 25q \cdot x - 0.72 \frac{x}{(1-x)}\frac{E}{q^2}\right) \qquad (12)$$

In (12) under the sign of the delta functions the values measured in the electron volts are placed, $\varepsilon_i(q) = A_i(q) - 2.4\gamma$, $A_i(q)$ represent the border of the bands, measured from the zero value of the crystal potential. Given that $0.72\frac{E}{1.2\gamma} = 0.6\frac{E}{\gamma} = 0.6 \cdot 0.5 MeV = 300000 eV$ the equation (12) can be rewritten as:

$$\frac{1}{12.5}\delta\left(A_i(q) - A_f(q(1-x)) - 2q \cdot x - \frac{x}{(1-x)}\frac{24000}{q}\right), \tag{13}$$

where $q = \frac{2EV_0}{\hbar^2 c^2 G_x^2} \rightarrow \frac{1}{4^2}\frac{2EV_0}{\hbar^2 c^2 G^2} = \frac{1}{4^2}\frac{2mEV_0}{\hbar^2 mc^2 G^2} = \frac{1}{4^2}\gamma\frac{2mV_0}{\hbar^2 G^2} = \frac{1}{4^2}\gamma V_0 / \frac{\hbar^2 G^2}{2m_0} \approx 8\gamma/16 = \frac{\gamma}{2}$

$$\tilde{a} = E_\perp^2 / 4G^2\hbar^2 c^2 - \frac{8EV_0}{2\hbar^2 c^2 G^2}; q = \frac{EV_G}{2\hbar^2 c^2 G^2}; \quad S = G_x x; \quad G_x \equiv nG = nG_{\text{мин}} = n2\pi/d.$$

Solve the equation for determining the spectrum of the emitted photons numerically. Take into account the fact that the energy spectrum of the CP both before and after emission of a photon meets the different bases. The Fig.5 shows the behavior of the delta function (13) depending on the emitted photon energy expressed in dimensionless fraction x of the incident fast particle energy in the case when the initial state's number coincides with the number of the final state band. The intersection of the curve with the ordinate axis of the graph corresponds to the vanishing of the argument of the delta function, ie the implementation of the laws of conservation of energy and momentum in the emission of a hard photon by the fast oriented particle. The photon energy x in the process of the photon emission can be found as a segment of the value of the intercept at the intersection of the curve with the horizontal axis. We call this value of x the possible energy of the emitted photon. During the emission of hard photons the CP band spectra of the initial i and final f states may have a fundamentally different nature in spite of the possible coincidence of the numbering of the bands. Impossibility of the emission of the high energy photons in non-dipole transitions between the states of the same band number of the above barrier particle is clearly seen from the Fig.5.

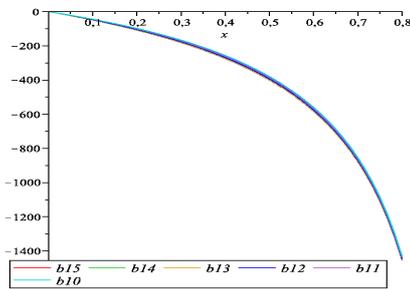

Fig. 5. Behaviour of the argument of the delta function (13) depending on the energy of the emitted photon x for the fast particle with q = 71 at the coincidence of the band numbers of the initial and final states. The curves for five above-barrier zones numbered from 10 to 15 practically coincide.

This conclusion follows from the fact that the potential energy of the emitted photon, as can be seen from Fig.5, is x = 0. The possibility of the emission of the high energy photons in non-dipole transitions between non-adjacent bands with different energies of the above-barrier particle is shown in Fig.6-8:

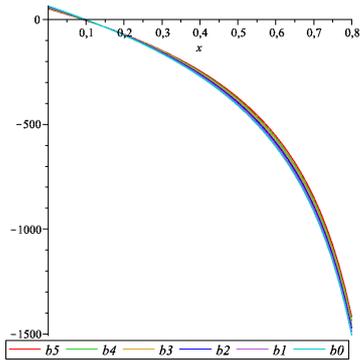

Fig. 6. Behaviour of the argument of the delta function (13) depending on the energy of the emitted photon x by the fast particle with q = 71 when changing CP band's number in the transition from the initial to the final state by two.

The curves for five overbarrier bands numbered from 10 to 15 are practically coinsiding. Emission of the high energy photons in non-dipole transitions between non-adjacent bands of the above-barrier particle is shown on the figs. 7,8. The initial set of bands and the final set of bands correspond to the different energies.

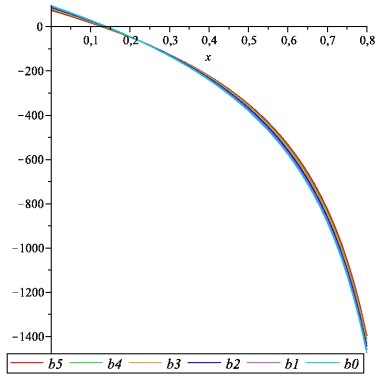

Fig.7. The argument of the delta function (13) behavior depending on the emitted photon energy x for the fast particle with q = 71 for the first six zones with increasing band's number in the quantum transition from the initial state to the final state by three.

The curves for the initial six above-barrier zones numbered from 0 to 6 have almost the same character. Emission of the high energy photons in non-dipole transitions between adjacent (nearest) bands of the above-barrier particle leads to the same conclusion. The Fig. 8 shows the same effect for the other CP energy.

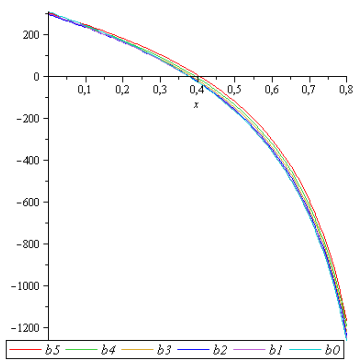

Fig.8. The argument of the delta function (13) behavior depending on the emitted photon energy x for the fast particle with q = 31 for the first six zones with increasing the numbers of the initial band when changing the band's number in the quantum transition from the initial state to the final state by two.

Weak dependence of the photon energy on the band number is clearly seen on the Figs 5-8.

**4. Orientation dependence of the probability of the band population.** Fig. 9 shows the occupancy ratios for odd and even zones for the case E = 28MeV, Si (110), depending on the angle of incidence of a positron in a crystal. Orientation dependence of the probability of the population of the lower boundary of the band for the positron in the even and odd states, respectively, depending on the angle of the incidence of the positron in a crystal relative to the plane (110) in the Si is shown. The angle θ is measured in the number of reciprocal lattice vectors corresponding to the projection of the momentum of the positron channeling across the planes, divided by the total momentum of the positron. Levels of oriented particles are numbered with the index bd. On the x-axis the wave vector is specified as a fraction of the reciprocal lattice vector. The four vectors of the reciprocal lattice at energy 28 MeV correspond approximately to an Lindhardt angle of the incidence of the positron, so that the value of the argument along the x-axis, which approximately equals to four, shares sub barrier and over barrier states.

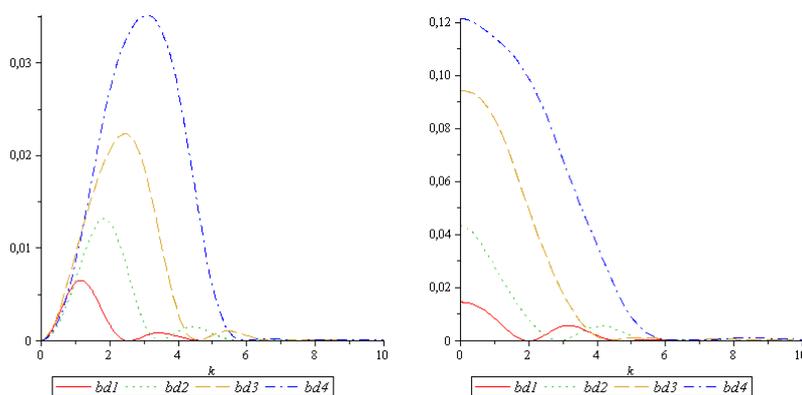

Fig. 9. Squared modulus of the matrix element of the transition of positron to the even and odd levels depending on the angle of incidence of the positron in a crystal. Angle θ is measured in reciprocal lattice vectors.

Levels of the oriented particle are numbered with the index bd. On the x-axis the wave vector is specified as a fraction of the reciprocal lattice vector. Five reciprocal lattice vectors at energy 28 MeV correspond approximately to an Lindhard angle of the incidence of the positron. Orientation dependence of the probability of the population of the positron's even and odd levels of the transverse motion depending on the *high* values of the angle of incidence of the positron in a crystal relative to the plane (110) in the Si is separately shown on the fig. 10.

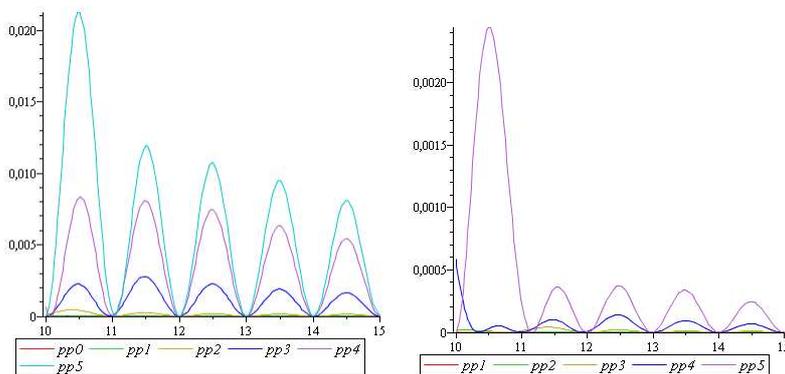

Fig. 10. Squared modulus of the matrix element of the transition of positron to the even and odd levels depending on the high values of the angle of incidence of the positron in a crystal.

The maximum population of near-barrier bands can be achieved by running the crystal positron beam at an angle to the crystallographic planes of the order of the Lindhardt angle . The Fig. 9 shows that at zero angles of incidence the deep sub-barrier bands are populated mostly, which in this case is usually interpreted as levels of the transverse motion, and at angles of incidence equal to the Lindhardt angle, mostly near-barrier states are populated.

**5. Matrix elements.** Numerical analysis shows that the highest probability of the CP quantum transition results from the matrix elements of quantum transition of the second kind $I^{(2)} = \langle \psi_n \exp(ikx) \partial/\partial x \, \psi_{n'} \rangle$ from (9). Modulus squares of the non-dipole complex matrix elements of the second kind between the states corresponding to the different energies of the CP are shown on the fig. 11.

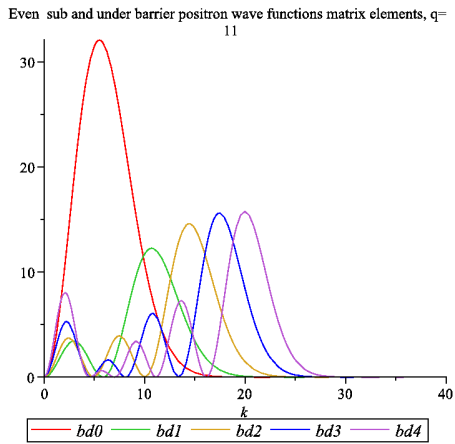

Fig. 11. Squared modulus of the non-dipole matrix elements of the second type for the transition of positron with q=71 to the energy q=51 between the even levels with the changing the number n of the CP transverse energy per unit depending on the high values of the angle of incidence of the positron on a crystal.

The calculated matrix elements of the first type are shown on the Fig. 12.

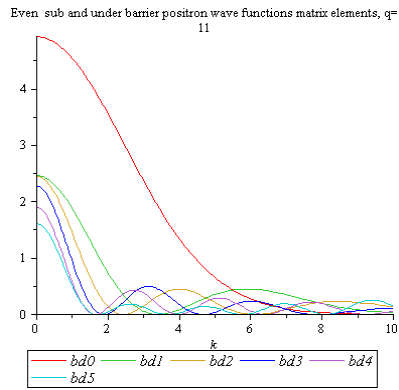

Fig. 12. Squared modulus of the non-dipole matrix elements of the first type for the transition of of positron with q=71 to the energy q=51 between the even levels with the changing the number n of the CP transverse energy per unit depending on the high values of the angle of incidence of the positron on a crystal.

The comparison of Fig. 11 and Fig. 12 shows that in the case of large pulses transmitted by the CP the matrix elements of the second kind are making the predominant contribution to the CP transition probability.

**6. Probability of nondipole radiation of photons.** Calculate the spectral-angular radiation density of the probability of the nondipole radiation of hard photons by the high-energy channeled particle with the energy $E$ which satisfies the inequality $U_0 E \geq 1$. It is known that non-dipole processes of the emission of hard photons can not be described at these energies in the quasi-classical approximation [1-9]. For unpolarized radiation the following expression for such a density of the probability can be written down [8]:

$$\frac{d^2 w(\omega,\theta)}{d\omega d\Omega} = \frac{e^2 \omega}{2\pi} \sum_f \left\{ \left(1+u+u^2/2\right)\left[\left|I_{if}^{(1)}(k_x)\right|^2 \theta^2 + \left|I_{if}^{(2)}(k_x)\right|^2 - 2\mathrm{Re}\left(I_{if}^{(1)}(k_x) I_{if}^{(2)*}(k_x)\right)\theta \cos\varphi \right] + \right.$$
$$\left. + \frac{u^2}{2E^2}\left|I_{if}^{(1)}(k_x)\right|^2 \right\} \times \delta\left\{ 2\frac{\omega}{\mathrm{E}-\omega}\left[\left(\theta^2 + \mathrm{E}^{-2}\right)\mathrm{E} - \omega\theta^2 \cos^2\varphi\right] - \varepsilon_i(\mathrm{E}) + \varepsilon_f(\mathrm{E}-\omega) \right\} \qquad (14)$$

As one can see from Fig.12, the quantum transition matrix elements of the channeled particle have significant values when the dimensionless component of the momentum of the channeled particle, directed across the crystal plane is significantly less than unity. This momentum in dimensionless form is measured in fractions of a reciprocal lattice vector which is equal to $2\pi/a = 2\pi \cdot 10^{10}$ m$^{-1}$ for the case of the silicon crystal. At the same time, the wave vector of the photon with the energy E of 10 GeV is equal to $k_{phot} = \hbar \mathrm{E}/c\hbar = = 10^{17}$ m$^{-1}$. Thus, for $k_x$ to have significantly lower values compared to the reciprocal lattice vector $k_x \leq 0.1 K$ the angle $\theta = k_x / k_{phot}$ should have values not exceeding $\theta \leq 2\pi \cdot 10^9 / 10^{17} \approx 6 \cdot 10^{-8}$. At the same time, the angle $\varphi$ in (14) can vary over a wide range, so that we can put $\delta\varphi \sim 0.1$. Consider the probability of the photon radiation by the channeled particle with the energy $\mathrm{E} = 10 GeV$ within a solid angle $d\Omega = \sin\theta d\theta d\varphi \approx \theta d\theta d\varphi \approx \approx 6 \cdot 10^{-8} \cdot 6 \cdot 10^{-8} \cdot 10^{-1} = 3.6 \cdot 10^{-16} sr$. Consider for example the even-even quantum transitions of the channeled particle when $\left|I_{if}^{(2)}(k_x=0)\right|^2 = 0$ (Fig. 11). For unpolarized radiation at the small observation angles $\theta \ll 1$, $\varphi \ll 1$ ($k_x \ll 1$) and in the limit $\theta \ll 1/E$, $\theta \ll u/E$, the expression (14) is reduced at the conditions specified before to the following simplified expression for the probability of the emission of the photon with the energy $\omega \leq E$ by the channeling particle:

$$\frac{d^2 w(\omega,\theta)}{d\omega d\Omega} = \frac{e^2 \omega}{2\pi} \sum_f \left(1+u_{if}+u_{if}^2/2\right)\frac{u_{if}^2}{2E^2}\left|I_{if}^{(1)}(0)\right|^2 \delta\left(\varepsilon_i(\mathrm{E}) - \varepsilon_f(\mathrm{E}-\omega) - \frac{\omega}{\mathrm{E}-\omega}\frac{2}{\mathrm{E}}\right), \qquad (15)$$

Here $u_{if} = \omega_{if}/(\mathrm{E}-\omega_{if})$ at the conditions specified by the nondipole radiation. Separating only one term in (15) it is easy to obtain the following probability of the photon emission with the photon energy $\omega_{if} = \mathrm{E}(\mathrm{E}-\omega_{if})\left[\varepsilon_i(\mathrm{E}) - \varepsilon_f(\mathrm{E}-\omega)\right]/2$:

$$\frac{dw_{if}(\theta)}{d\Omega} = \int d\omega \frac{d^2 w_{if}(\omega,\theta)}{d\omega d\Omega} = \frac{e^2 \omega_{if}}{\pi} \frac{u_{if}^2}{4E^2}\left(1+u_{if}+u_{if}^2/2\right)\left|I_{if}^{(1)}(0)\right|^2 \qquad (16)$$

Consider $\hbar\omega_{if} \approx 0.4\mathrm{E}$ so that $u_{if} \sim 0.66$. At the conditions of the nondipole radiation specified before $\hbar\omega_{if}/\hbar = 0.4 \cdot 10^{10} eV / 0.66 \cdot 10^{-15} eV \cdot s = 0.6 \cdot 10^{25} s^{-1}$. Turning to the dimensional units and taking into account the fact that $\left|I_{if}^{(1)}(k_x \approx 0)\right|^2 \approx 5$ (Fig.12) obtain the following estimate for the transition probability:

$$w_{if}(\omega_{if}) = dw_{if}(\omega_{if})d\Omega = \frac{e^2}{\hbar c \pi}\frac{\hbar\omega_{if}}{\hbar}\frac{(\hbar\omega_{if})^2}{(E-\hbar\omega_{if})^2}\frac{(mc^2)^2}{E^2}(1+u_{if}+u_{if}^2/2)\left|I_{if}^{(1)}(k_x\approx 0)\right|^2 d\Omega = \quad (17)$$

$$= 2\cdot 10^{-3}\cdot 0.6\cdot 10^{25}\cdot(0.66)^2\cdot(4\cdot 10^4)^{-2}1.5\cdot 5\cdot 3.6\cdot 10^{-16}\cdot 1/s \approx 0.2\ s^{-1}.$$

Differential with respect to the dimensionless frequency of the emitted photon $x=\omega/E$ probability $\frac{dw(\omega,\theta\ll 1)}{d\omega}$ for the generation of photons by a channeled particle which entered the crystal at an Lindhard angle is shown on the Fig.13:

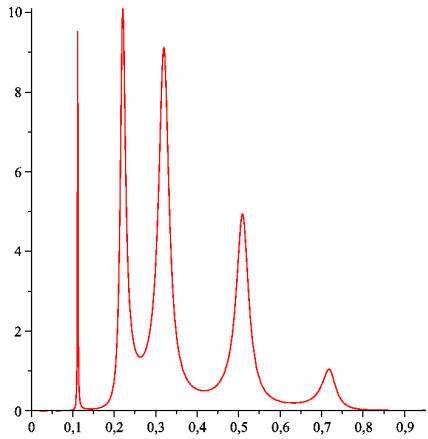

Fig.13. Differential probability $\frac{dw(\omega,\theta\ll 1)}{d\omega}$ (s.$^{-1}$) for the generation of photons by a channeled particle with q=71. On the x-axis dimensionless photon energy x as the part of the energy of the channeled particle is postponed.

**7. Resume.** The calculation of the quasi-Bloch energy spectrum of the oriented fast charged particle entering the crystal at an angle both greater and smaller than the Lindhard angle is performed. It is shown that the band structure with the presence of allowed and forbidden bands has been preserved during the passage of fast charged particles above the crystal potential. The processes of the photon generation by the quantum crystal-oriented particle entering into the crystal at an angles both greater and smaller than the Lindhard angle are considered. The probability of the photon excitation by the quantum above-barrier channeled particle is calculated. It is shown the predominant role of the matrix elements of the second-type for the transition between the quantum levels of the transverse motion of the CP. It is shown that the emission spectrum of hard photons consists of a set of the well-observed emission lines. The nondipole photon emission processes by the fast charged oriented particle have been considered. The above-barrier and sub-barrier transitions as well as quantum bands of the above-barrier and sub-barrier channeled particle have been considered on an equal footing. The parametric dependence of the band properties of channeled particle on the particle energy is fully taken into account . The probability of the nondipole photon radiation by the channeled particle with the energy $E=10\,GeV$ within a solid angle $3.6\cdot 10^{-16}\,sr$ oriented along the propagation direction of the channeling particle is shown to be equal to $w_{if}(\omega_{if})=0.2$ s$^{-1}$. When extending the solid angle within which the measurements are made, the probability of the photon emission increases significantly. Such an increase in the probability is associated with the fact that at large angles of observation the predominant contribution to the probability of the photon radiation will bring the transition matrix elements of the second type $\left|I_{if}^{(2)}(k_x)\right|^2$, that are much larger in magnitude than the matrix elements of the first type $\left|I_{if}^{(1)}(k_x)\right|^2$ (Figs. 11,12). It has been shown that the nondipole processes of the radiation of hard photons by the channeled particles are well observed experimentally.